\documentclass[12pt]{article}
\pdfoutput=1
\usepackage{epsfig}
\usepackage{amsfonts}
\usepackage{amssymb}
\usepackage{xcolor}
\usepackage{amsmath}
\usepackage{upgreek}
\usepackage{graphicx}
\usepackage{cite}
\usepackage{calligra}
\usepackage{caption}
\usepackage{subcaption}

\newcommand{\nc}{\newcommand}
\nc{\beq}{\begin{equation}} \nc{\eeq}{\end{equation}}
\nc{\beqa}{\begin{eqnarray}} \nc{\eeqa}{\end{eqnarray}}
\nc{\ba}{\begin{array}} \nc{\ea}{\end{array}}

\begin{document}
\begin{center}

{\bf \LARGE Effective potential in leading logarithmic approximation in non-renormalisable $SO(N)$ scalar field theories} \vspace{1.0cm}

{\bf \large  R.M. Iakhibbaev$^{1}$ and D. M. Tolkachev$^{1,2}$} \vspace{0.5cm}

{\it $^1$Bogoliubov Laboratory of Theoretical Physics, Joint Institute for Nuclear Research, 
  6, Joliot Curie, Dubna, Russia\\and \\
$^2$Stepanov Institute of Physics,
68, Nezavisimosti Ave., Minsk, Belarus}
\vspace{0.5cm}

\abstract{The study of the effective potential for non-renormalisable scalar $SO(N)$ symmetric theories leads to recurrence relations for the coefficients of the leading logarithms. These relations can be transformed into generalised renormalization-group (RG) equation which can be analyzed in detail. In some special cases this equation can be solved exactly.}
\end{center}

\section{Introduction}

The effective potential plays a very important role in the determination of vacuum properties in weakly coupled field theory, which was pointed out in the pioneering work of Coleman and Weinberg \cite{CW}. The calculation of this object by summing infinite series of Feynman diagrams with zero external momentum is a laborious task even for renormalisable interactions. Fortunately, in the leading logarithmic approximation one can easily find an exact expression for the effective potential due to renormalisability of potentials well studied in the literature \cite{CW, EA,  Chung:1997jy, Kastening:1996nj}. For non-renormalisable theories, even the very statement of a such a problem seems at first glance infeasible due to the fact that the structure of the counterterms differs from the structure of the original non-renormalisable Lagrangian, quantum corrections to the Lagrangian grow uncontrollably.

Not so long ago there has been developed a technique for studying the leading logarithms of the effective potential $V_{eff}$ in scalar theories with non-renormalisable interactions. In the paper \cite{Kazakov:2022pkc}, a scalar potential of arbitrary form was studied, which has no any symmetries. The formalism of the paper \cite{Kazakov:2022pkc} was based on the extraction of information about the leading logarithmic approximation from vacuum Feynman diagrams in the framework of the Jackiw approach \cite{EA}, and on the application of the $\cal{R}$-operation and the Bogoliubov-Parasiuk theorem \cite{BP,Hepp,Zimmermann} on loop Feynman graphs. Thus, it is possible to find a relation between the leading UV-singularities of $n$-loop vacuum diagrams and leading poles of $(n-1)$-loop diagrams. Finally, one can figure out that higher divergences are governed by one-loop diagrams, as in the renormalisable case.  This fact leads to recurrence relations that generalize the standard renormalization-group (RG) equations.

In this work we try to generalize the analysis to a four dimensional scalar theory with $SO(N)$-type interactions, this Lagrangian can be represented in a generalized form as
\beq
{\cal{L}}=\frac{1}{2}\partial_\mu \phi_a \partial^\mu \phi_a-g V(\phi_a \phi_a).
\eeq
For convenience of the analysis we restrict ourselves to the power and exponential potentials, which in general can be represented as
\begin{equation}
    g V(\phi_a \phi_a)= \frac{g}{p!}(\phi_a \phi_a)^{p/2},
\end{equation}
where $p \geq 4$ is an integer number,  or
\begin{equation}
   g V(\phi_a \phi_a)= g \exp(|\phi_a|/m) 
\end{equation}
where $a=1,2,\ldots,N$, $m$ is the dimensional parameter and $|\phi_a|$ is the length of the vector $\phi_a$. In the general case, these potentials are non-renormalisable, as seen from the power counting.

The structure of the paper is as follows. In the first part of the paper, we briefly discuss the derivation of the effective potential within the functional formalism and the Feynman rules derived from the shifted action. In the second part of the paper, we calculate the first loop orders contributing to the effective potential. In the second section, we also briefly discuss the $\mathcal{R}'$-operation and how recurrence relations can be obtained in connection with the Bogoliubov-Parasiuk theorem.  Finally, we analyze the behaviour of the obtained recurrence relation in different limits and explore the generalised RG-equation in the large $N$-limit.

\section{Effective potential and vacuum diagrams}

The generating function of the considered theory is given as \cite{Buchbinder:2021wzv} 
\beq
Z(J)=\int \mathcal{D} \phi ~ \exp \left(i\int d^4x ~ {\cal L}(\phi,d \phi)+J\phi \right),
\eeq
thus $Z(J)$ corresponds to vacuum-to-vacuum transitions in the presence of the classical external current $J(x)$. It is known that the effective potential is obtained by the Legendre transform of the connected functional $W(J)=-i\log Z(J)$ 
\beq
\Gamma(\phi)= W(J)-\int d^4x J(x)\phi(x),
 \eeq
where the classical field $\phi(x)$ is defined as the solution to $\phi(x)=\frac{\delta W(J)}{\delta J(x)}$. However, it is convenient to directly read the effective potential in a perturbative sense through 1PI Feynman graphs; so the effective potential can be represented as
\beq
V_{eff}(\phi)=g\sum_{k=0}^\infty (-g)^k V_k(\phi).
\label{effpotV}
\eeq
The rules for the latter can be obtained from the shifted action $S[\phi + \hat{\phi}]$ where $\phi$ is the classical field obeying the equation of motion and $\hat {\phi}(x)$ is the quantum field over which integration can be performed~\cite{EA}. 
This shift with $\phi_a$ being a scalar field with $N$-components gives rise to a mass matrix, which can be conveniently represented as \cite{EA,Kastening:1996nj,Chung:1997jy}
\beq
m_{ab}^2=\hat{v}_2 \left(\delta_{ab}-\frac{\phi_a\phi_b}{\phi^2}\right)+v_2 \frac{\phi_a \phi_b}{\phi^2} \label{massMatrix}
\eeq
where
$$m_1^2=gv_2=g\frac{\partial^2 V}{\partial \phi^2},\\~
m_2^2=g\hat{v}_2=2g \frac{\partial V}{\partial (\phi^2)}.$$

\begin{figure}[ht]
 \begin{center}
  \epsfxsize=10cm
 \epsffile{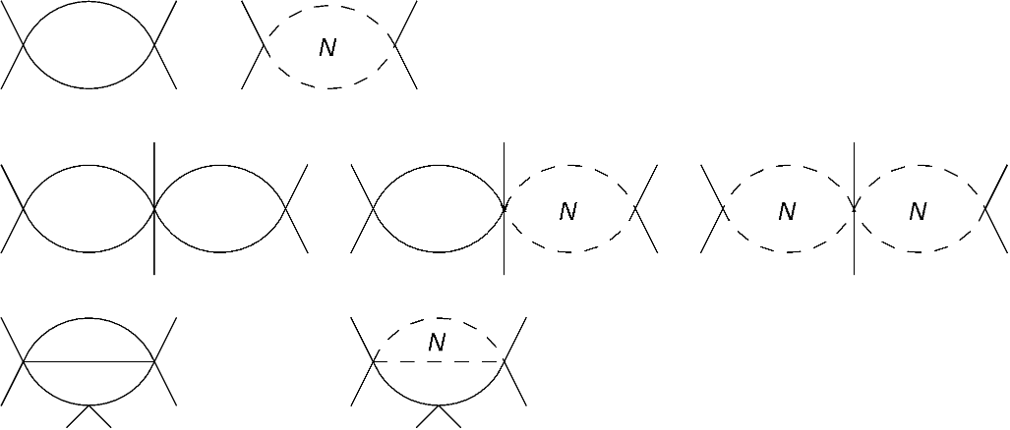}
 \end{center}
 \vspace{-0.2cm}
 \caption{Vacuum graphs contributing to the effective potential up to two loops} 
\label{vacgr}
 \end{figure}

One can note that the non-diagonal terms (that give contribution proportional to $\hat{N}=N-1$) and the diagonal terms are factorised (this technique was used in \cite{Chung:1997jy,Kastening:1996nj} to study higher loop corrections to the effective potential). According to this representation, one can write propagators in the following form
\begin{equation}
    G_{1,ab}=\frac{1}{p^2-m_1^2} \frac{\phi_a \phi_b}{\phi^2}, \label{mass1}
\end{equation}

\begin{equation}
    G_{2,ab}=\frac{1}{p^2-m_2^2} \left(\delta_{ab}- \frac{\phi_a \phi_b}{\phi^2}\right). \label{mass2}
\end{equation}
The vacuum diagrams contributing to the effective potential are depicted in Fig. \ref{vacgr}.
The vertices are also generated by shifting and expanding the corrections (we denote them as $\Delta V_1$) by the quantum field $\hat{\phi}$ \cite{Kazakov:2022pkc}. In our case the vertices correspond to the derivative of the classical $SO(N)$-potential: 
\beq
v_n=\frac{\partial^n V}{\partial \phi^n}.
\eeq
Here the derivative corresponds to the number of internal lines without the vector $SO(N)$ index. Also, the shift gives vector vertices with indices as follows 
\beq
\hat{v}_2=2 \frac{\partial V}{\partial (\phi^2)}\;,  \; \hat{v}_{a,3}=2 \frac{\partial}{\partial \phi_a} \frac{\partial V}{\partial (\phi^2)} ~\text{and also} \;  \hat{v}_{3}=2 \frac{\partial}{\partial \phi} \frac{\partial V}{\partial (\phi^2)}
\eeq
and so on. For the triple vertex with two internal lines carrying indices we have 
\beq
\Lambda_{a b c}=(\delta_{a b} \hat{v}_{c,3}+\delta_{a c} \hat{v}_{b,3}+\delta_{b c} \hat{v}_{a,3}).
\eeq
For the internal lines of the quartic vertices carrying the vector index one must contract propagators with symmetric tensor:
\beq
X_{abcd}=\left(\delta_{ab} \delta_{cd}+\delta_{ac}\delta_{bd}+\delta_{ad}\delta_{bc}\right).
\eeq
The same rules hold for the higher loop contribution and more complicated vertices \cite{Chung:1997jy}.

The Feynman rules derived above allow us to compute loop diagrams (we use the dimensional regularization $d=4-2\epsilon$) for the effective potential so that the one-loop contributions to the effective potential have the following form
\begin{equation}
    \Delta V_1=\frac{v_2^2}{4}\left(\frac{1}{\epsilon}+\log\left(\frac{m_1^2}{\mu^2}\right)\right)+\hat{N}\frac{\hat{v}_2^2}{4}\left(\frac{1}{\epsilon}+\log\left(\frac{m_2^2}{\mu^2}\right)\right),
\end{equation}
where $1/4$ is the combinatorial factor and $\mu$ is the dimensional transmutation parameter.
Computation of two-loop diagrams gives the following expression for leading divergences:
\beq
\Delta V_2=\frac{1}{8\epsilon^2}\left(v_2 v_3^2+v_4 v_2^2\right)+\hat{N}(\hat{N}+2)\frac{\hat{v}_4 \hat{v}_2^2}{4\epsilon^2}+3\hat{N} \left(\frac{v_{3}^2 \hat{v}_2}{\epsilon^2}+\frac{\hat{v}_{3}^2 v_2}{\epsilon^2}\right)
 \eeq 
Using these expressions, one can get all the expressions obtained earlier \cite{Chung:1997jy}. For example, at the one-loop level of leading-log contribution for $(\phi^2)^2$-potential one can get 
\begin{equation}
   V_1= \frac{1}{4} \frac{g (\phi^2)^2}{4} \log \left(\frac{m_1^2}{ \mu^2}\right)+\frac{\Tilde{N}}{4} \frac{g  (\phi^2)^2}{36} \log \left(\frac{m_2^2}{\mu^2}\right) \label{oneloopf4}
\end{equation}
and at the two-loop level
\begin{equation}
    V_2= \frac{3 g^2 (\phi^2)^2}{32} \log^2 \left(\frac{m_1^2}{ \mu^2}\right)+\Tilde{N}\frac{g^2(\phi^2)^2}{48} \log \left(\frac{m_1^2}{\mu^2}\right) \log\left(\frac{m_2^2}{ \mu^2}\right)+\Tilde{N}^2\frac{g^2 (\phi^2)^2}{864} \log^2 \left(\frac{m_2^2}{ \mu^2}\right)\label{twoloopf4}
\end{equation}
To obtain higher order contributions, we turn to the BPHZ procedure which allows us to relate the lower orders in the PT-series to the higher orders. 

\section{BPHZ-procedure and RG-equation}

\begin{figure}[h!]
 \begin{center}
  \epsfxsize=13cm
 \epsffile{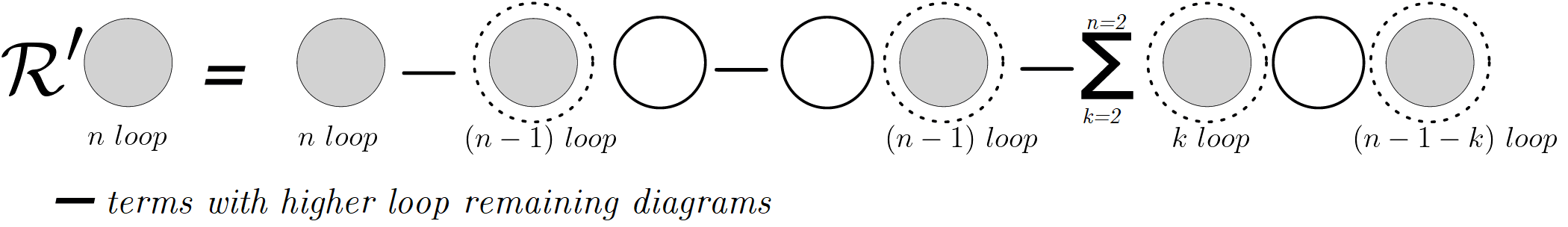}
 \end{center}
 \vspace{-0.2cm}
 \caption{$\cal{R}'$-operation applied to $n$-loop divergent diagram} 
\label{rprime}
 \end{figure}

Let us remind that the ${\cal R}$-operation~\cite{BogoliubovBook}  being applied to an $n$-loop diagram subtracts first of all the ultraviolet divergences in subgraphs starting from one loop up to $(n-1)$- loops and then finally subtracts the remaining $n$-loop divergence, which are obliged to be local due to the Bogoliubov-Parasyuk theorem~\cite{BP,Hepp,Zimmermann}. This $n$-loop divergence left after the incomplete ${\mathcal{R}}$-operation ($\mathcal{R}'$-operation) is precisely what we are looking for.
The locality requirement tells us that the leading $1/\epsilon^n$ divergence in $n$-loops $A^{(n)}_n$ is given by the following
formula~\cite{we2015}:
\beq
A^{(n)}_n=(-1)^{n+1}\frac 1n A^{(1)}_n, \label{red}
\eeq
where $A^{(1)}_n$ is the one-loop divergence left after subtraction of the $(n-1)$-loop counter term as a result of the incomplete ${\cal R'}$-operation. Recall that the $\mathcal{R'}$-operation for any graph $G$ can be defined  recursively via the action of the $\mathcal{R'}$-operation on divergent subgraphs~\cite{Vasiliev, Collins}:
\begin{equation}
    \mathcal{R}'G= \left(1- \sum_\gamma K\mathcal{R}'_\gamma+\sum_{\gamma \gamma' } K\mathcal{R}'_\gamma K\mathcal{R}'_{\gamma'} - \ldots  \right) G,
\end{equation}
where the subtraction operator $K_\gamma$ subtracts the UV divergence of a given subgraph $\gamma$. The action of the $\cal{R}'$-operation is shown in Figure \ref{rprime}

Using this theorem and applying it in the same way as in \cite{Kazakov:2022pkc}, one can write the expression for the generalized renormalization equation for non-renormalisable theories. The recurrent structure of the leading divergences can be written as
\beq
n \Delta V_n=\frac{1}{4} \sum_{k=0}^{n-1} D_{ab} \Delta V_{k} D_{ab}V_{ n- k-1}\label{RG}
\eeq
with the differential operators
$$
D_{ab} \to \frac{\partial^2}{\partial \phi_a \partial \phi_b}
$$
This equation can be re-expressed 
\beq
n \Delta V_n= \frac{1}{4} \sum_{k=0}^{n-1} \left(4 \hat{N} \frac{\partial}{\partial (\phi^2)} \Delta V_k \frac{\partial}{\partial (\phi^2)} \Delta V_{n-k-1}+\frac{\partial^2}{\partial \phi^2} \Delta V_k\frac{\partial^2}{\partial \phi^2}  \Delta V_{n-k-1}\right), \ \ \ n\geq 2 \label{rec}
\eeq
This equation can be considered as the main result of the present work. If we choose $N=1$, we see that we recover the generalized renormalization-group equation in the case of ordinary arbitrary scalar interaction \cite{Kazakov:2022pkc}. Obviously, no analytical solution can be obtained from the general equation since even the trivial limit $N=1$ does not yield an equation that can be solved. However, it can be seen that in the limit of large $N$ the recurrence equation is considerably simplified. It can be shown that if one represents the recurrence relation in the large $N$ limit (which in itself means summation over all bubble-type topologies)
\begin{equation}
    n \Delta V_n= \sum_{k=0}^{n-1} \left(\hat{N} \frac{\partial}{\partial (\phi^2)} \Delta V_k \frac{\partial}{\partial (\phi^2)} \Delta V_{n-k-1}\right).
\end{equation}
It can be rewritten as a differential equation after introducing $$\Sigma=\sum_{n=0}^\infty (-z)^n \Delta V_n (\phi)$$
as
\begin{equation} \label{RGEN}
   \boxed{ \frac{d}{dz}\Sigma(z,\phi)=-N\left(\frac{\partial}{\partial \phi^2} \Sigma(z,\phi)\right)^2, ~\Sigma(0,\phi)=V_0} 
\end{equation}

The effective potential can be represented as a function of the solution with the interchanged variable
\begin{equation}
    V_{eff}=g \Sigma(z,\phi)\big|_{z \rightarrow -\frac{g}{16 \pi^2}\log(g\hat{v}_2/\mu^2 )} \label{varchangelog}
\end{equation}
Note that the RG-equation \eqref{RGEN} is an ordinary differential equation of the first order and can be solved analytically at least for a number of cases of non-renormalisable potentials. It can be seen that the generalised renormalisation group equations look roughly the same as in the renormalisable case with the exception that the beta function is replaced by a complex operator of fields. This situation fits quite well into the general picture of the study of non-renormalisable theories \cite{Kazakov:2020xbo}.

Let us turn to particular examples of potentials to which one can calculate quantum corrections with the help of the obtained generalised renormalisation-group equation \eqref{RGEN}. 

\section{Large $N$ limit}

\subsection{Power-like potentials}

Now let us consider an example of a power-like potential. 
It is convenient to obtain a solution to the generalised RG-equation in the form of the following ansatz:
\begin{equation}
    \Sigma(z,\phi_i)=\frac{(\phi^2)^{p/2}}{p!} f(z (\phi^2)^{p/2-2}) \label{ansatzp}
\end{equation}
where $p$ is a power of initial interaction.
This behaviour is justified by the fact that the expansion of the function $f(z)$ is given as a power series, as one can see in (\ref{oneloopf4}~-\ref{twoloopf4}). Inserting \eqref{ansatzp} into \eqref{RGEN} one easily can find an ordinary differential equation with the initial conditions
\begin{equation}
   - \frac{N}{4p!} \left((p-4) x f'(x)+p f(x)\right)^2=f'(x), ~f(0)=1
\end{equation}
where the dimensionless argument is introduced as it comes from loop expansion $x=z (\phi^2)^{p/2-2}$ . One can notice that this equation is homogeneous (we can use it and look for solution as $f(Nx)$).

\begin{figure}%
    \centering
    \subfloat[\centering $f(z)$ behaviour in the renormalisable case]{{\includegraphics[width=7cm]{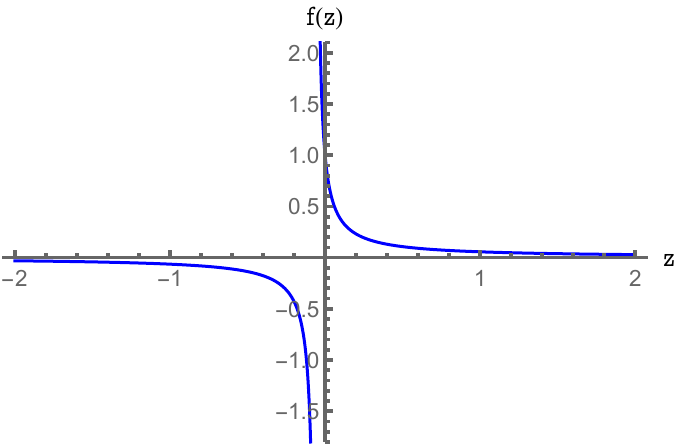} }}
    \qquad
    \subfloat[\centering  normalized effective potential $\textbf{Re} (V_{eff})$ with $\mu \sim 1$ and $g \sim 1$.]{{\includegraphics[width=7cm]{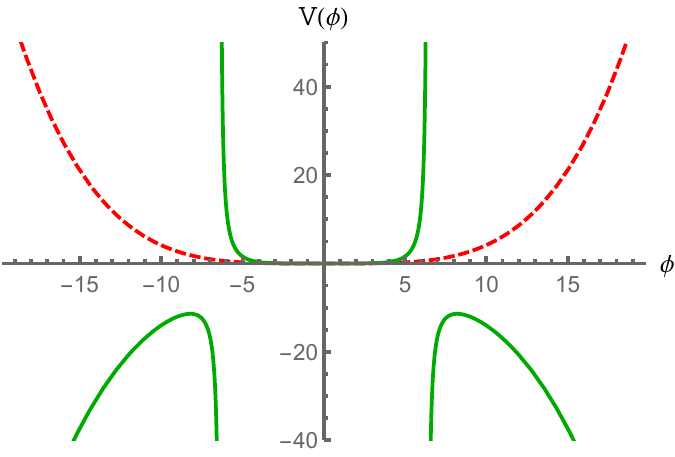} }}
    \caption{Effective potential in the renormalisable case for $N=100$. The red dashed line is for the classical quartic  $SO(N)$-potential, and the green thick line is for RG-improved effective potential}%
    \label{figph4}%
\end{figure}

In the simplest $p=4$ case, the equation turns into sum of a geometric progression as expected:
\begin{equation}
    f(z)= \frac{1}{1 + \frac{N}{6} z},
\end{equation}
and here the pole is located at
\begin{equation}
     z=-6/N.
\end{equation}
Thus, in the renormalisable case the generalised RG-equation \eqref{RGEN} reproduces the well-known textbook example. It is easy to change variables as prescribed in \eqref{varchangelog} and get a full expression for the effective potential
\eqref{effpotV}:
\begin{equation}
    V_{eff}=\frac{g(\phi^2)^2/4!}{1-\frac{N}{6}\frac{g}{16 \pi^2}\log(\frac{m_2}{\mu})}.
\end{equation}
The corresponding graphics for $f(z)$ and $V_{eff}$ are shown in Figure \ref{figph4}. In these figures one can clearly see a pole due to which the ground state in such renormalisable theory is stable. 

Let us turn to non-renormalisable examples (when $p>4$). It is easier to shift the ansatz
\begin{equation}
    f(x) \rightarrow ((4-p) x)^{\frac{p}{4-p}} S(x)
\end{equation}
and to find that
\begin{equation}
    p S(x)-(p-4) x S'(x)=\frac{N}{4p!}  ((4-p) x)^{2\frac{p-6}{p-4}}S'(x)^2, ~ S(0)=0.
    \label{genrgeqdef}
\end{equation}
Thus, nonlinearity is simplified and isolated. Unfortunately, the general solution of the equation in analytically closed form cannot be obtained, but it should be noted that the solution of the equation by quadrature is given by
\begin{equation}
   -\frac{2 (p!)^{\frac{p-4}{2}} \left(-2 \sqrt{\frac{N (p-4) }{(p-1)!}q S(x)+1}-p+2\right)^{\frac{p-2}{2}} \left(\sqrt{\frac{N (p-4)}{(p-1)!} q S(x)+1}-1\right)}{N (p-4) p^2 q S(x)}=C.
\end{equation}
Here $C$ is the constant of integration and $q=((4-p) x)^{\frac{4}{4-p}}$, hence the desired function in exact form must be expressed as cumbersome inverse function.

\begin{figure}%
    \centering
    \subfloat[\centering $f(z)$ behaviour in the $(\phi^2)^3$-case]{{\includegraphics[width=7cm]{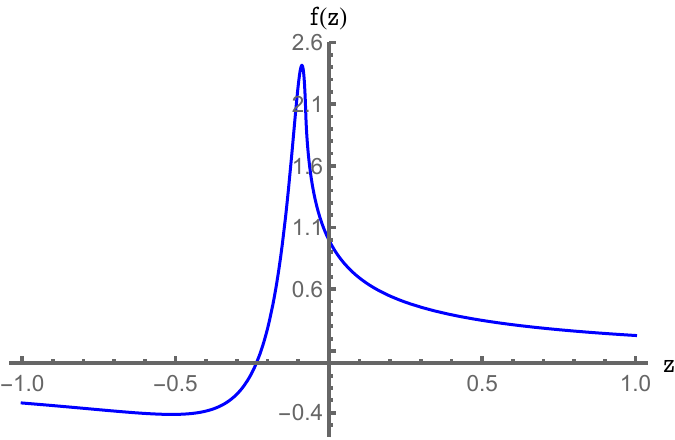} }}
    \qquad
    \subfloat[\centering normalized effective potential $\textbf{Re} (V_{eff})$ with $\mu \sim 1$ and $g \sim 1$.]{{\includegraphics[width=7cm]{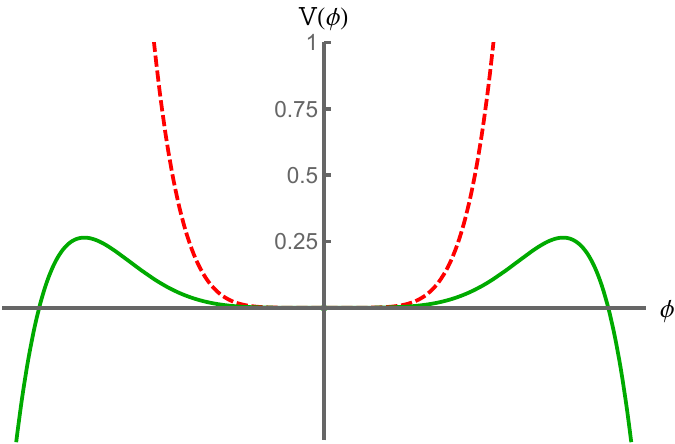} }}
    \caption{Effective potential in the $(\phi^2)^3$ case for $N=100$. Red dashed line is for classical sextic  $SO(N)$-symmetric potential, green thick line is for RG-improved effective potential}%
    \label{figph6}%
\end{figure}

In some cases one can solve this generalised RG-equation in an analytically exact way. The most obvious and easiest choice of potential is $p=6$. In this case, we can get from \eqref{genrgeqdef}
\begin{equation}
    -\frac{N}{720} \left(x S'(x)+3 S(x)\right)^2=S'(x)
\end{equation}
and solution can be found as
\begin{equation}
    f(x)=3\times 5!^2\frac{ \left(\frac{1}{2} N x \left(1+\frac{N x}{5!}\right)-10 \left(1+\frac{N x}{30}\right)^{3/2}+10\right)}{N^3 x^3}\label{largeNphi6}
\end{equation}
Note that this expression describes a number of bubble diagrams including those generated by the vertices of $\phi^6$-type.  Note that in  \eqref{largeNphi6} there is no pole or function discontinuity behaviour. It is easy to see that \eqref{largeNphi6} has an imaginary part when
$$
x<-30/N.
$$
However, the behaviour of the function remains regular as shown in Fig. \ref{figph6}. It is known that imaginary parts (due to non-convexity of solution \eqref{largeNphi6}) have a natural explanation as indicators of unstable states \cite{Weinberg:1987vp}.

\subsection{Exponential potential}

\begin{figure}%
    \centering
    \subfloat[\centering $f(z)$ behaviour in the exponential case]{{\includegraphics[width=7cm]{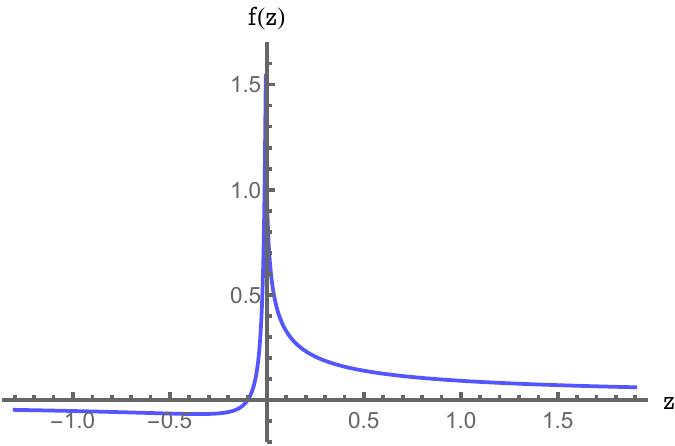} }}
    \qquad
    \subfloat[\centering normalized effective potential $\textbf{Re} (V_{eff})$ with $\mu \sim 1$ and $g \sim 1$.]{{\includegraphics[width=7cm]{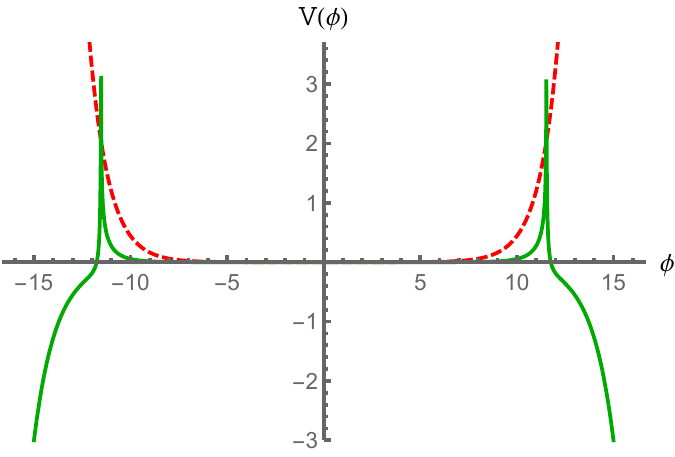} }}
    \caption{Effective potential in the $e^{|\phi|/m}$ case for $N=100$. The red dashed line is for the classical exponential  $SO(N)$-symmetric potential and the green thick line is for the RG-improved effective potential}%
    \label{figexp}%
\end{figure}

It is quite interesting that representation of the mass matrix in the form of \eqref{massMatrix} is also applicable to the exponential potential. To proceed, we can perform the same steps for the $\exp(|\phi|/m)$-potential and look for solution of RG-equation in the following representation
\begin{equation}
    \Sigma(z, \phi_a)=e^{|\phi|/m} f(x)
\end{equation}
where we introduced dimensionless variable $x=z/m^4 e^{|\phi|/m}$. Thus the RG-equation has the following form
\begin{equation}
   \frac{N}{4} \left(x f'(x)+f(x)\right)^2=-f'(x) \label{expLargeN}
\end{equation}
with the initial condition
\begin{equation}
    f(0)=1 \label{inicondexp}
\end{equation}
Solution of \eqref{expLargeN} satisfying \eqref{inicondexp} is given by
\begin{equation}
   f(x)=\frac{1}{2 N x}W( N x) (W(N x)+2)
\end{equation}
where $W(x)$ is the Lambert function. This solution is regular, $f(x)$ is presented in Fig. \ref{figexp}. Again, there are no signs of irregularity but the appearance of the imaginary part (the exact value of the argument is given by the transcendental equation $\textbf{Im}(f(x))=0$). Surprisingly, the resulting expression is an analytic function of $x$ although the original potential is not. It should be noted that here the initial potential is included as a variable in the argument of the function so that in fact the function is really not analytic. Thus, as in the previous example, the ground state of the obtained solution can be considered as metastable.

\section{Conclusion}

In this paper, we have succeeded in constructing a generalised renormalisation-group equation for an arbitrary scalar theory with $SO(N)$ symmetry using the Bogoliubov-Parasiuk theorem. The resulting equation reproduces in the limit the equation we studied earlier. In the large $N$ limit we managed to obtain a generalised renormalisation-group equation which turn out to be solvable. In some special cases we showed its exact solutions and the corresponding effective potentials. The common property of the resulting effective potentials in non-renormalisable models is their metastability and absence of behavioural peculiarities i.e. discontinuities.

The exactly solvable equations obtained allow the properties of effective potentials to be described more qualitatively even in the case of non-renormalisable interactions. Therefore, there is a good prospect of studying more complicated actions with more interesting properties, for example,  four-dimensional supersymmetric actions of the Wess-Zumino type and their generalisations (as is well-known, kahlerian and chiral superpotentials can be supplemented by quantum corrections \cite{Huq,Buchbinder:1994iw}). Also, effective potentials are quite interesting from the point of view of studying the scheme dependence since effective potentials require the computation of only quite simple vacuum diagrams. Thus, the study of the sub-leading logarithmic order promises to be fruitful as it was in the case of higher dimensional on-shell four-point scattering MHV-amplitudes  \cite{Kazakov:2019heg,we2017}.

\section*{Acknowledgments}
The authors are grateful to D.I. Kazakov for reading the manuscript and valuable comments. The authors would also like to thank I.L. Buchbinder and S.V. Mikhaylov for fruitful discussions.

\bibliographystyle{unsrt}
\bibliography{refs}
\end{document}